\documentclass[aps,prl,reprint, amsmath,amssymb,author-year]{revtex4-1}

\usepackage{graphicx}
\usepackage{dcolumn}
\usepackage{bm}

\begin{document}

\title{An all-optical Compton source for single-exposure X-ray imaging}
\author{A. D\"opp$^{1,2}$, E. Guillaume$^1$, C. Thaury$^1$, J. Gautier$^1$, I. Andriyash$^1$, A. Lifschitz$^1$, J-P. Goddet$^1$, A. Tafzi$^1$, V. Malka$^1$, A. Rousse$^1$ and K. Ta Phuoc$^1$}
\address{$^1$ LOA, ENSTA ParisTech - CNRS - Ecole Polytechnique - Université Paris-Saclay, 828 bd des Maréchaux, 91762 Palaiseau Cedex France}
\address{$^2$ Centro de Laseres Pulsados, Parque Cientfico, 37185 Villamayor, Salamanca, Spain}

\begin{abstract}
\noindent 

All-optical Compton sources are innovative, compact devices to produce high energy femtosecond X-rays. Here we present results on a single-pulse scheme that uses a plasma mirror to reflect the drive beam of a laser plasma accelerator and to make it collide with the highly-relativistic electrons in its wake. The accelerator is operated in the self-injection regime, producing quasi-monoenergetic electron beams of around 150 MeV peak energy. Scattering with the intense femtosecond laser pulse leads to the emission of a collimated high energy photon beam. Using continuum-attenuation filters we measure significant signal content beyond 100 keV and with simulations we estimate a peak photon energy of around 500 keV. The source divergence is about 13 mrad and the pointing stability is 7 mrad.  We demonstrate that the photon yield from the source is sufficiently high to illuminate a centimeter-size sample placed 90 centimeters behind the source, thus obtaining radiographs in a single shot.
\end{abstract}


\maketitle

X-ray radiography is one of the most important non-invasive diagnostic tools used in medicine and industry. Conventional sources are primarily based on two processes that convert electron energy to high energy photon radiation: bremsstrahlung and synchrotron radiation \cite{Jackson:1990uq}. Synchrotron sources can provide collimated, polarized, low bandwidth radiation \cite{Bilderback:2005fw}, yet X-ray tubes using the simpler bremsstrahlung conversion are by far more common as source of X-rays. This is due to the substantial capital costs and dimensions of synchrotron light sources, which typically consist of a GeV-level electron accelerator, a large scale storage ring and beam lines with insertion devices to produce radiation.

Laser-based light sources have the potential to produce synchrotron-type radiation in drastically reduced physical size and at a reasonable cost. In modern laser-plasma accelerators (LPAs) \cite{Faure:2004tj,Geddes:2004vs,Mangles:2004vr,Esarey:2009ks} electrons can reach GeV energies over centimeter-scale acceleration lengths \cite{Leemans:2006ux}. In addition to their compactness, such sources have a competitive normalized transverse emittance, due to their micrometer source size, and provide inherently ultrashort, femtosecond duration beams \cite{Lundh:2011js}. Furthermore, using an optical undulator instead of conventional magnetic structures can reduce the undulator period $\lambda_0$ by 3-4 orders of magnitude. Hence the ensemble of laser-driven accelerators and undulators permits to reproduce the principle of a synchrotron even on a millimeter scale.

The radiation produced in an optical undulator is similar to conventional synchrotron radiation. As discussed in \cite{Corde:2013bja}, the peak deflection angle of highly-relativistic electrons during the interaction with an electromagnetic wave depends on the intensity and wavelength. Notably the undulator parameter $K$ in this configuration is given by the normalized peak vector potential $a_0\simeq  0.85 \lambda [\mu\mbox{m}]\sqrt{I[10^{18}\mbox{W/cm}^2]}$. But in contrast to conventional sources, such an inverse Compton source has an additional degree of freedom, which is the scattering angle $\phi$ between laser and electrons. The fundamental wavelength of backscattered photons is then given by \cite{Ride:1995vh}
$$\lambda = \frac{\lambda_0}{2\gamma^2(1-\beta\cos\phi)}\left( 1+\frac{a_0^2}{2}+\gamma^2\theta^2\right).$$
Here $\theta$ denotes the observation angle, while $\beta=v/c_0$ and $\gamma=(1-\beta^2)^{-1/2}$ are the normalized electron velocity and the Lorentz factor, respectively. The source is most efficient in a counter-propagating configuration, upshifting the backscattered radiation by up to $4\gamma^2$. For typical short-pulse laser parameters (e.g. $hc_0/\lambda_0\approx1.55$ eV for a Ti:Sa laser with $\lambda_0=800$ nm) this means that electrons exceeding 200 MeV can produce MeV-level photons.

The combination of LPAs with backscattering is an experimentally challenging task, which usually requires extensive control of two terawatt-class laser pulses. In particular the electron beam and the scattering laser have to overlap in time and space, as was demonstrated for example in \cite{Powers:2013bx,Khrennikov:2015gx}. To avoid the alignment and synchronization issues of this two-beam scheme, we use a different method that is based on the retro-reflection of a single laser pulse \cite{TaPhuoc:2012cg}. Here the intense drive pulse of the LPA is used for Compton scattering with the electrons traveling in its wake, as well. To achieve this, the beam is backreflected by a plasma mirror, which is placed close to the exit of the LPA. The advantage of such scheme is that it inherently ensures an overlap of the electron beam and the laser pulse. In addition, it operates close to 180 degree, which permits the best conversion efficiency to high energy photons. While the first proof-of-principle experiment \cite{TaPhuoc:2012cg} showed a very broad photon energy distribution, recent studies have observed that scattering with quasi-monoenergetic electron beams can produce X-rays with about 50 percent energy spread at full width at half maximum (FWHM) \cite{Tsai:2015uo}. \newpage

\onecolumngrid

\begin {figure}[t]
\centering
\includegraphics[width=0.99\linewidth]{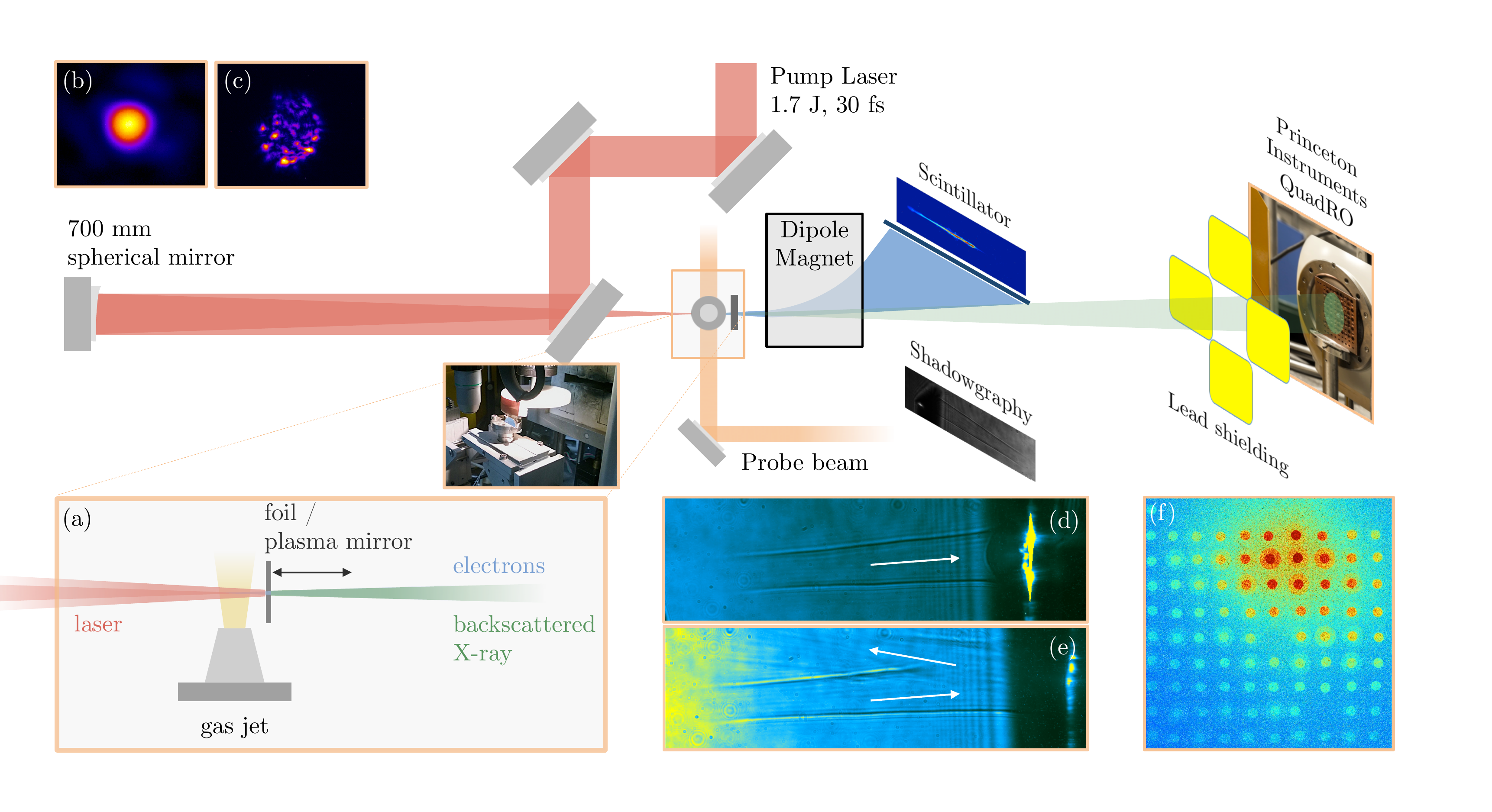}
\caption{Setup of the Compton-backscattering experiment. The foil which acts as plasma mirror is placed at the exit of the gas jet, as shown in (a). The foil can be moved along the axis to distinguish between bremsstrahlung and backscattering. (b) shows the focal spot at the gas jet entrance. There are two diagnostics for the backreflection: (c) shows an image of the backscattered light on the foil, which is recollected using a lens system. Though not absolutely calibrated, this measures the relative intensity of the backscattered light. Also, when there is no reflection we observe a clear plasma channel on the shadowgraphy (d). Once the foil is in place we can also see the channel created from the reflected beam (e). The emitted X-rays are detected downstream on a Gadox scintillator, fiber-coupled to a low noise CCD (Princeton Instruments QuadRO). The temperature of the radiation is estimated using a filter array, as shown in (f).}
\label{Fig1}

\end{figure}

\twocolumngrid

Here we show that the X-ray signal of such backscattered beams is sufficient to make radiographs of macroscopic samples in a single shot. The experiment was performed with the \textsc{Salle Jaune} Ti:Sa laser system at Laboratoire d'Optique Appliqu\'ee, which delivers linearly polarized laser pulses with 30 femtosecond duration (FWHM). Using a spherical mirror of 700 mm focal length the pulse is focussed at the entrance of a supersonic helium gas jet of 2 mm diameter, see Fig.\ref{Fig1}. The focal spot contained 50-55$\%$ of the beam energy, delivering an energy of about 0.9 J on target.

\begin {figure}[t]
\centering
\includegraphics[width=1.0\linewidth]{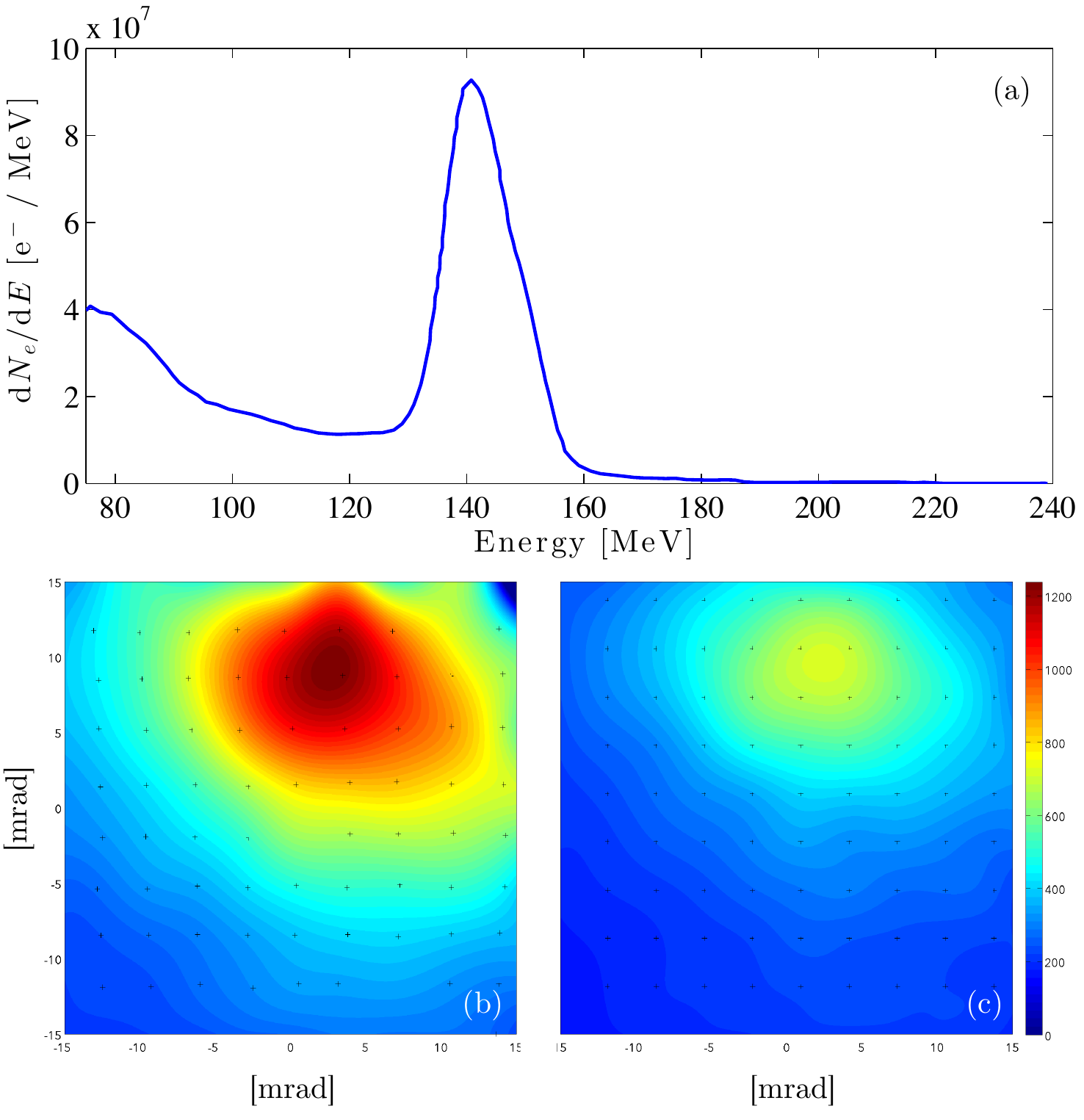}
\caption{Typical electron beam spectrum used in the experiment (a). The spectrometer cuts electrons with energies below 75 MeV. Lower part: Reconstructed signal of Compton backscattered radiation without (b) and with a 5.1 mm copper filter (c).}
\end{figure}

The laser-plasma accelerator was operated in the transverse self-injection regime, producing beams of 300 - 600 pC charge at beam energies mostly in the range of 100 - 150 MeV. Electron charge and spectra were measured using a dipole magnet spectrometer combined with an absolutely calibrated phosphor screen. In contrast to previous experiments \cite{TaPhuoc:2012cg} the electron beams in this study had higher charge and showed important quasi-monoenergetic features, cf. Fig.2a. 

For the plasma mirror a 100 $\mu$m cellophane foil was used. This choice of the plasma mirror material aims to reduce the production of bremsstrahlung as electrons pass through the plasma mirror. The foil was slightly inclined with respect to the laser axis in order to avoid backreflection into the laser chain. It also permitted to observe the plasma channel created by the reflected beam in the gas jet, see Fig.1e. Furthermore the reflected laser beam was imaged on a screen. After each shot the foil was translated to provide an undamaged mirror surface. 

X-rays were detected using a GdOS based scintillator, fiber-coupled to a 16-bit CCD (Princeton Instruments Quad-RO 4320). The total detector area of the camera is $5\times5$ cm$^2$, divided into $2084\times 2084$ pixels of 24 $\mu$m edge length each. Placed on the laser axis at 90 centimeters from the interaction, this lead to a field of view of approximately $55\times 55$ mrad$^2$.

The efficiency of the production of Compton scattering radiation depends on the intensity of the backreflected laser at the position of collision with the electrons. It therefore depends on the position of the foil with respect to the gas jet exit. As the Rayleigh length of the laser is about 0.8 millimeters, any signal observed for the foil placed at one centimeter behind the gas jet is not due to Compton backscattering but a result of bremsstrahlung produced by electrons passing through the foil. This background signal is rather stable, with a relative rms error of $\sigma_{brems}\approx 0.1$. When moving the foil inwards we observe a signal increase, up to about $3.2\pm0.6$ times the bremsstrahlung background at the optimal foil position at the gas jet exit. 

\begin {figure}[b]
\centering
\includegraphics[trim=1cm 0cm 1cm 1cm, clip=true,width=1.0\linewidth]{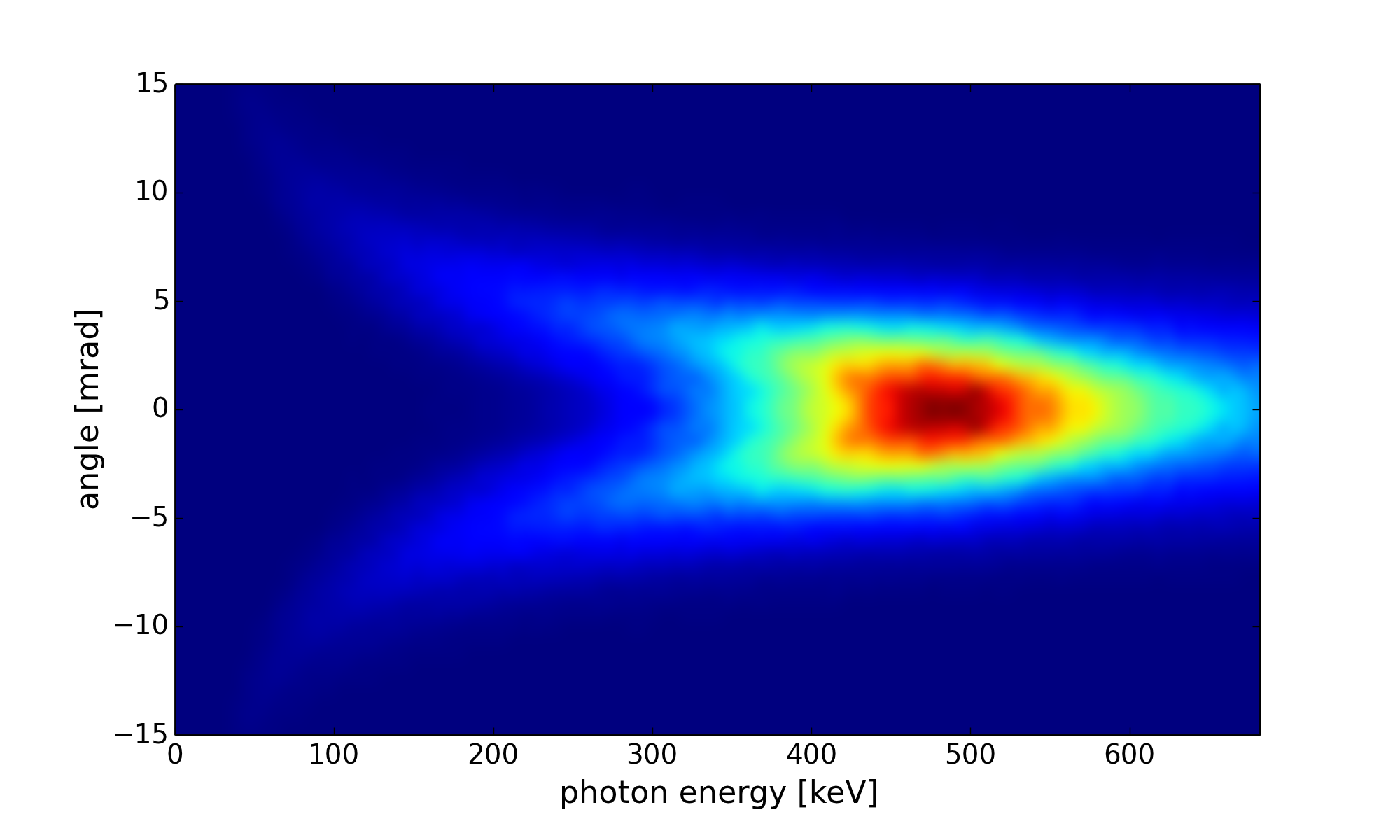}
\caption{Simulated angularly resolved spectrum of backscattered photons from the interaction of a laser-accelerated electron beam (140 MeV, 5 \% energy spread, 5 mrad divergence) with a high power laser pulse  ($a_0=1.0$, 30 fs).}
\label{Xspectrum}
\end{figure}

Additionally, we could distinguish the signals due to their very different spectra. For this, we placed a 5.1 mm copper filter in front of the detector (shown in Fig.1f), which absorbs around 99 (90) percent of radiation below 70 (100) keV. Holes inside the filter plate permit reconstruction of the unperturbed signal level, which is shown in Fig.2b. The ratio between both the filtered signal $I_1$ and the signal that passes through the holes $I_0$ allows to roughly estimate the spectral content. We observed an important difference between the beams from Bremsstrahlung radiation and Compton backscattering. Shots with the foil close to the jet showed a high contrast ratio of 0.6, the contrast being defined as $1-I_{1}/I_{0}$. When moving the foil outwards the jet the contrast reduced to 0.1. This implies that the Compton X-rays are much more attenuated and therefore contain less signal beyond 100 keV than the Bremsstrahlung beam (which extends far into the MeV range). The large field of view permitted us to measure the beam divergence. We observed a FWHM beam divergence of $\Theta_x=(12.7\pm 3.6)$ mrad and $\Theta_y=(13.0\pm 4.0)$ mrad, where $x$ is plane of laser polarization, with a pointing stability of 8.3 mrad and 6.4 mrad, respectively. While not directly measured, the duration of the pulse is approximately given by the electron bunch duration \cite{Corde:2013bja}, which is typically a few femtoseconds for self-injected electron beams. 

\begin {figure}[b]
\centering
\includegraphics[width=.85\linewidth]{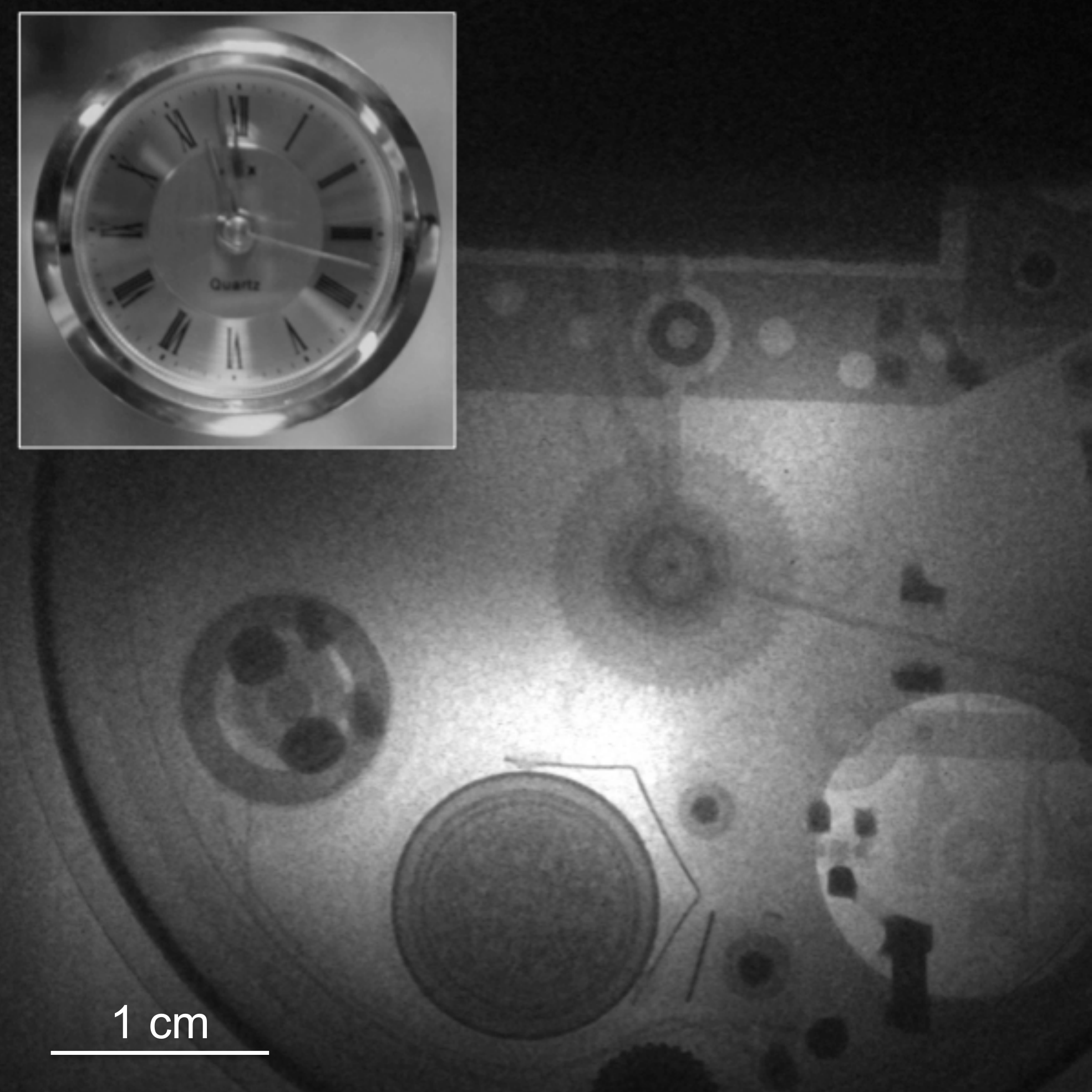}
\caption{Single-exposure X-ray image of a clock with a photography as inset. Different materials are clearly identifiable. The dark area in the upper part is a shadow cast by the dipole magnet employed to measure the electron spectrum. The signal values measured in this area are used to estimate the background signal and outliers have been removed.}
\end{figure}

To get a better estimate of the X-ray spectrum, we simulated the interaction between the electron bunch and the laser in a test particle model using \textsc{PlaRes} \cite{Andriyash:2015th}. For these simulations we used 5000 test particles to sample an electron beam of 140 MeV peak energy, 5 percent energy spread and 5 mrad divergence, similar to the electron beam distribution shown in Fig.2a. Not included in this model is the dark current, which will contribute to the low energy tail of the spectrum. We modeled the  scattering pulse as a counter-propagating laser pulse with peak potential $a_0=1.0$ and 30 fs duration. For such beam parameters the backscattered radiation fields add up incoherently \cite{Corde:2013bja} and as shown in Fig.3 we observe that the resulting radiation spectrum is much broader than the initial electron spectrum. The large energy spread is a consequence of the elevated undulator strength parameter $a_0=1.0$, the initial electron energy spread and the electron beam divergence. Simulations for various beam parameters show that the latter is usually the dominating effect for laser-accelerated electron beams. Taking into account the missing dark current contribution in the simulation and the reduced scintillator response at higher photon energies, our results are in fair agreement with the experimental data.

To demonstrate that the X-ray flux from Compton backscattering is sufficient for imaging applications, we performed radiographies of macroscopic objects, placed in front of the detector. As an example Fig.4 shows the radiography of a clock. Its structural parts consist of different metal alloys, while the gears are made of plastic. The different materials are clearly distinguishable in the radiography due to their distinct absorption coefficients. As discussed previously the bremsstrahlung component is very energetic ($>100$ keV) and therefore adds to the image mostly in form of a subtractable background noise. Due to the inherent small source size of the source \cite{TaPhuoc:2012cg} a subject placed closer to the source would show significant edge enhancement, as used for propagation-based phase contrast imaging \cite{Bravin:2012hp}. In this regard the Compton source is similar to, but more energetic than, the laser-based betatron source \cite{Najmudin:2014uu}.

In conclusion we have performed experiments on Compton-backscattering of an intense laser with electrons from a laser-plasma accelerator. The micrometer-size optical undulator period leads to the emission of radiation of hundreds of keV, well beyond energies obtained with other femtosecond sources as for example betatron sources or free-electron lasers \cite{Corde:2013bja}. Laser wakefield based Compton scattering therefore provides unique properties that may be of interest for ultrafast pump-probe experiments, e.g. for initial fusion research. Beyond applications in fundamental research, the source has also promising properties for medical and industrial applications. In particular we have shown that the photon flux of a single shot is sufficient for imaging of macroscopic objects. Due to the inherent micrometer source size the source is suitable for phase contrast imaging techniques, which will be an area of future investigations. \\

\footnotesize{ACKNOWLEDGMENTS: We acknowledge the Agence Nationale pour la Recherche
through the FENICS Project No. ANR-12-JS04-0004-01 and the FEMTOMAT Project No. ANR-13-BS04-0002, the LA3NET project
(GA-ITN-2011-289191), and the GARC project 15-03118S.}

\bibliography{radiography.bib}

\end{document}